\documentclass[conference]{IEEEtran}
\IEEEoverridecommandlockouts
\usepackage{cite}
\usepackage{amsmath,amssymb,amsfonts}
\usepackage{algorithmic}
\usepackage{graphicx}
\usepackage{textcomp}
\usepackage{xcolor}
\def\BibTeX{{\rm B\kern-.05em{\sc i\kern-.025em b}\kern-.08em
    T\kern-.1667em\lower.7ex\hbox{E}\kern-.125emX}}
\begin{document}

\title{Analysis of Comments Given in Documents Inspection in Software Development PBL and Investigation of the Impact on Students\\

}



\maketitle

\begin{abstract}
This study considers inspection conducted in software development PBL as learning feedback and investigates the impact of each inspection comment on students. The authors have already collected most inspection comments for not only requirements specification but also UML diagrams on GitHub.  The authors develop a tool in GitHub that collects comments given in Figma. We examine the impact on students of each classification of inspection comments based on the post-lesson questionnaire submitted by the students.
Finally, we present the benefits that classification of inspection comments can bring to PBL and discuss automatic comment classification by machine learning enabled by text-based comments and the concept of software development PBL support application enabled by automatic classification of inspection comments.

\end{abstract}


\section{Introduction}
Active learning that learners actively participate in their learning is paid attention in recent years.
Human resource development with high level of software development skills is also required.
From these background, project-based learning (PBL)for software development in higher education institutes is paid attention and numerous practices have been reported such as Sakulviriyakitkul, et ak. \cite{saku}，and Chuna, et al. \cite{paulo}.
The authors have been conducting PBL for software development for more than twenty years \cite{haze1}.
\par
This course conducts the waterfall-based software development in the form of project-based learning by novice software developers (university students).
The development process includes peer-review in the teams as well as inspection by the teaching staff (teacher and teaching assistants who passed the course) for the artifacts of the upstream phases.
\par
The original purpose of software inspection is quality assurance.
However, a study reported inspection has advantages on not only quality assurance but also internal audit, communications between departments and competency promotion \cite{retro}.
Based on the trend, we regard inspection as a feedback mechanism in the field of education.
\par
This study investigates how inspection affects learning of students. To achieve this goal, we develop a tool that transports inspection comments conducted outside of GitHub to GitHub \cite{git} and classify all the inspection comments into categories previously developed.
Then we discuss the trends between classification results and the results gained from questionnaire.
\par
Finally, we propose usage of classification results for the teacher and students and describe future directions of the automatic classification of inspection comments.

\section{Related work}
This section introduces related work of this study from the viewpoint of importance of feedback in education, feedback in software engineering education and classification of inspection comments.
\subsection{Importance of feedback in education}

We stand the position that feedback is important in education. Hattie et al. described that feedback brought strong impacts in learning and its outcomes \cite{feedback}. They also pointed out that effectiveness of feedback varied by the timing of feedback, and the types and way of it.

\subsection{Feedback in software engineering education}
We think the inspection process in software development plays one kind of feedback.
We introduce related work that focused on feedback in software engineering education.
Anvar et al\cite{review}. introduced peer review among students for evaluation of a large course of user-centered design because the teacher cannot evaluate all the students. They described the peer review functioned as feedback. However, they didn’t clarify what kind of feedback was given and how they affect students’ learning.

\subsection{Classification of inspection comments}
Gilb and Graham classified inspection comments based on document inspection in ThornEMI, an English company as the following categories \cite{gilb}:
\begin{itemize}
   \item shortage of data
   \item inaccurate data
   \item additional data
   \item inconsistent data
   \item other (standards etc.)
\end{itemize}
\par
Hazeyama classified comments of inspection conducted in his PBL course into thirteen categories\cite{haze2}. Table I shows the classification. As the classification by Hazeyama includes that of Gilb and Graham, this study adopts that of Hazeyama.
\begin{table}
  \centering
    \caption{comment classification of hazeyama}
    \begin{tabular}{|p{3 cm}|p{5 cm}|}  \hline
      classification & explanation \\ \hline \hline
      short description & Lack of description \\ \hline
      excess & Excess of description \\ \hline
      abstract &  Too abstract description \\ \hline
      understandability & Difficult to understand the descriptions \\ \hline
      undefined & Undefined term \\ \hline
      inconsistent & Inconsistency among deliverables \\ \hline
      mistake & Obvious errors in description and model notation \\ \hline      
      rationale & Unknown design basis \\ \hline
      short items & Lack of information to be stated \\ \hline
      missed inspection comments & Unmodified to previous inspections\\ \hline
      presentation  & Inappropriately worded \\ \hline
      enhancement request & Suggestions for improvements to specifications\\ \hline
      format & Document formatting deficiencies \\ \hline
    \end{tabular}
\end{table}

\section{Overview of PBL in practice}
This section describes an overview of the course, the development process, the artifacts required to create by the teams and inspection process. 
\subsection{Overview of the course}
The course is offered to third-year undergraduate students in the Department of Informatics Education at Tokyo Gakugei University, Japan. The team is organized into three to five students. The course consists of 15 weekly 90-min long lectures. The task given to the students is web application development using Java. In the preceding semester, we provide an introductory software engineering course. 
\par
The development process is based on the waterfall model. Verification activities, that is, software inspection and testing, are conducted. Software inspection is conducted by the teaching staff (teaching assistants and the teacher) for artifacts of the requirements specification, UML diagrams, and a database design document. After system testing by the teams, acceptance testing is conducted by the teaching staff.
\par
To provide feedback to the student teams, the inspection of artifacts created during the upstream phase, and acceptance testing of the application developed by each team are conducted by the teaching staff.

\subsection{Artifacts}
The types of artifacts that each team is required to create are requirements specification, a user interface design document, a class diagram, a database design document, sequence diagrams, source codes, unit/system testing reports, development plan, team progress reports (each week), and a project completion report as shown in Table II. The artifacts except for the user interface design document are written in text, therefore inspection comments are also written in text. As we will describe in section IV, we succeed in collecting inspection comments from Figma in text. We can deal with all the inspection comments in text.

\begin{table}
  \centering
    \caption{documents and explanation}
    \begin{tabular}{|c|p{5 cm}|}  \hline
      document & explanation \\ \hline \hline
    Functional specification & Describes the specific functions of the system(Markdown) \\ \hline
    Screen transition diagram & Description of system screen design and transition(figma) \\ \hline
      class diagram & Describes the data design class of the system(plant UML) \\ \hline
      Database specification & Describe the database design of the system(Markdown) \\ \hline
      Sequence Diagram & Describe specific sequences for each function(plant UML) \\ \hline
      state chart diagram & If there is a state change in the system, describe the state change.(plant UML) \\ \hline
  \end{tabular}
\end{table}

\subsection{Development environments}
As the development environment, students use their own laptop computers. We use GitHub \cite{git} as a source code and document repository. We use Figma \cite{figma} for creating user interface design document.

\subsubsection{GitHub}
We use the version control in the documents, the “Issues” function (the formal location of our text communications, including discussions and defect reporting, among other exchanges), and the “Pull Request” function for the artifact review process of GitHub. We have proposed an inspection process that issues the pull request from the branch the target artifact of inspection is put to the master branch \cite{miyashita}.
\subsubsection{figma}
Figma enables to design user interfaces and transitions of flows in a graphical user interface (GUI) manner. In addition, comments can be attached to arbitrary elements. These functions are satisfied with the requirements of our inspection process, therefore, we adopt Figma. 
\section{Comment aggregation across platforms}
In this study, I categorize all the inspections comments and consider their relationship to the students' learning. In the process, it is necessary to organize comments collected on different platforms, GitHub and Figma, in a centralized manner.
\par
Then we developed a Python script, SceneCommenter, which automatically submits Figma comments to GitHub and aggregates all PBL inspections comments in a GitHub repository.
\par
SceneCommenter, given a Figma project ID and a GitHub pull request ID, retrieves the text of every comment that exists in the Figma project, the coordinates of the comment, and the image file of the screen where the comment was recorded through FigmaAPI for each comment in the Figma project.
\par
The acquired screen image and coordinate information are used to create an image of a pin on the screen that indicates the location where the comment was posted. By generating this image, teachers can visually check the location of comments made with Figma on GitHub.
\par
Figure1 shows an example of a screen image after compositing.
This image is first stored in a repository for image storage. Using GitHubAPI, the newly created image is stored by retrieving Ref, retrieving Commit, creating Blob, creating Tree, creating Commit, and updating Reg.
\par
Finally, it posts a comment to the pull request. Using the GitHub API, it creates a comment including reference information to the uploaded image in the specified Pull Request. Figure 2 shows the flow of SceneCommenter operation, and Figure 3 shows the GitHub screen after it is activated.
\par
By developing this script, I was able to consolidate all the inspection comments handled in the class into GitHub. By retrieving all the comments via GitHub API, it is possible to organize the comments in csv format.
\begin{figure}
  \centering
  \includegraphics[width=\linewidth]{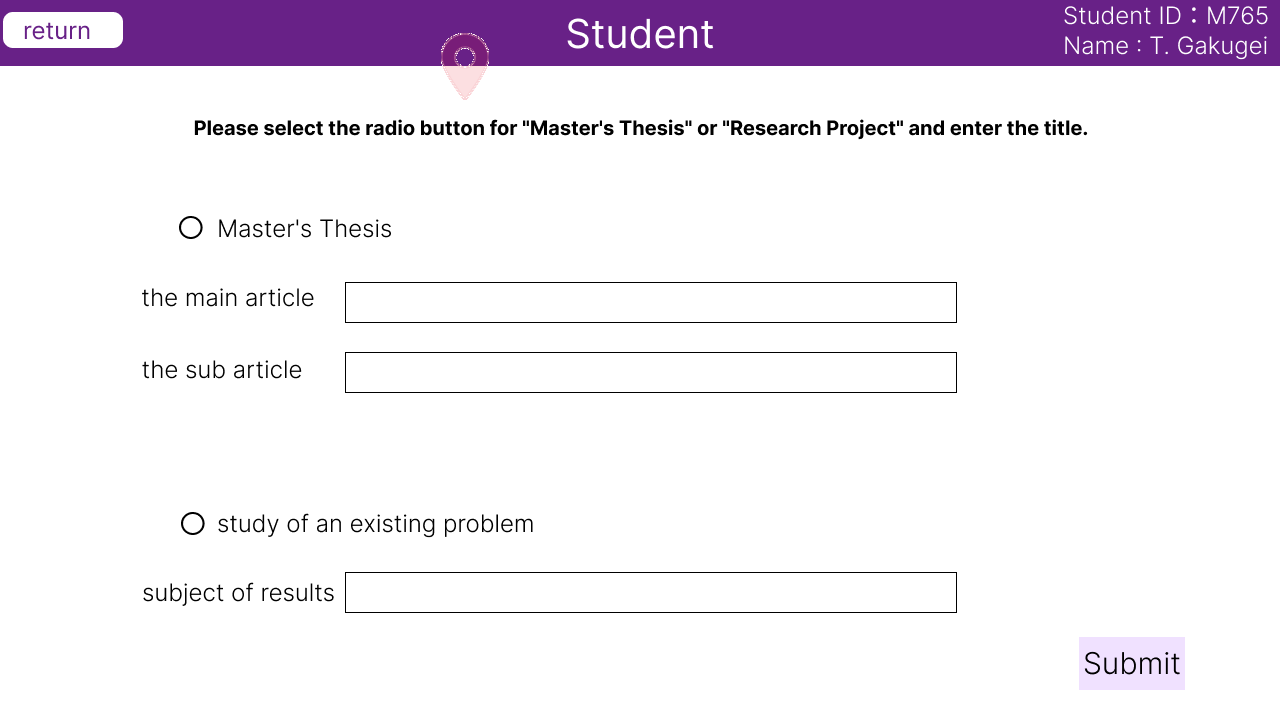}
  \caption{Image automatically generated based on Figma information}
\end{figure}
\begin{figure}
  \centering
  \includegraphics[width=\linewidth]{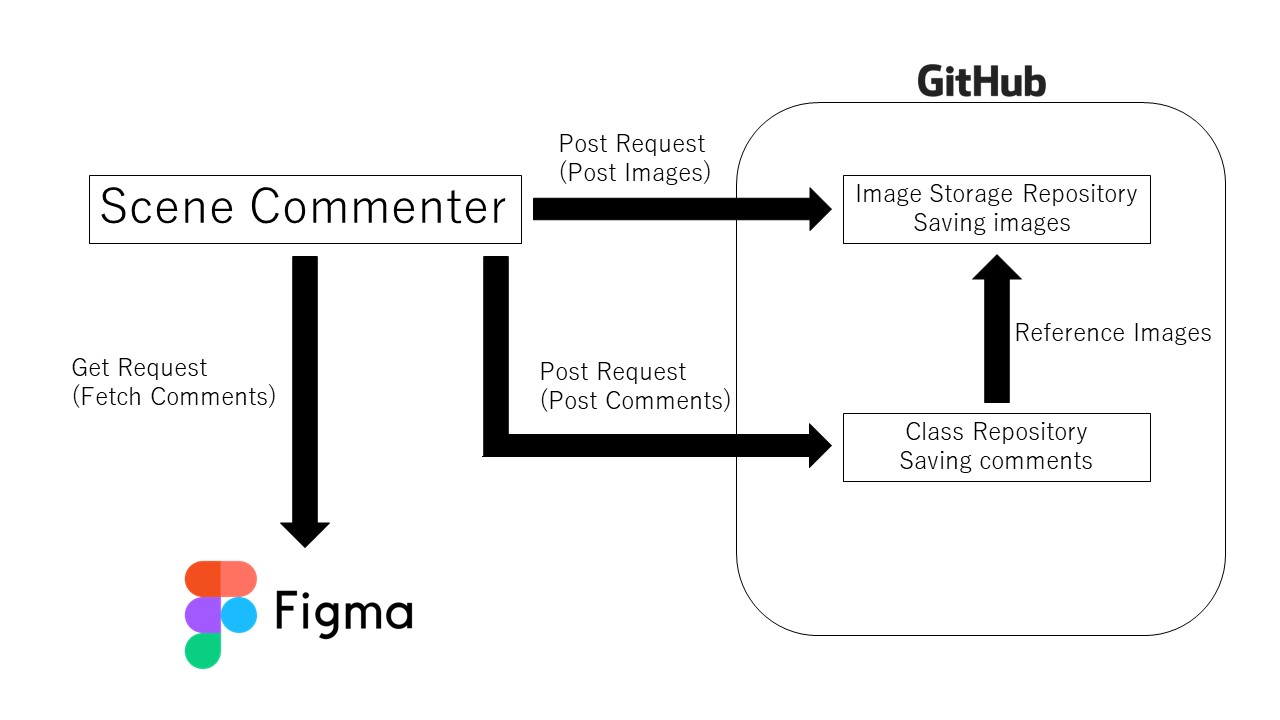}
  \caption{
Flow of SceneCommenter operation}
\end{figure}
\begin{figure}
  \centering
  \includegraphics[width=\linewidth]{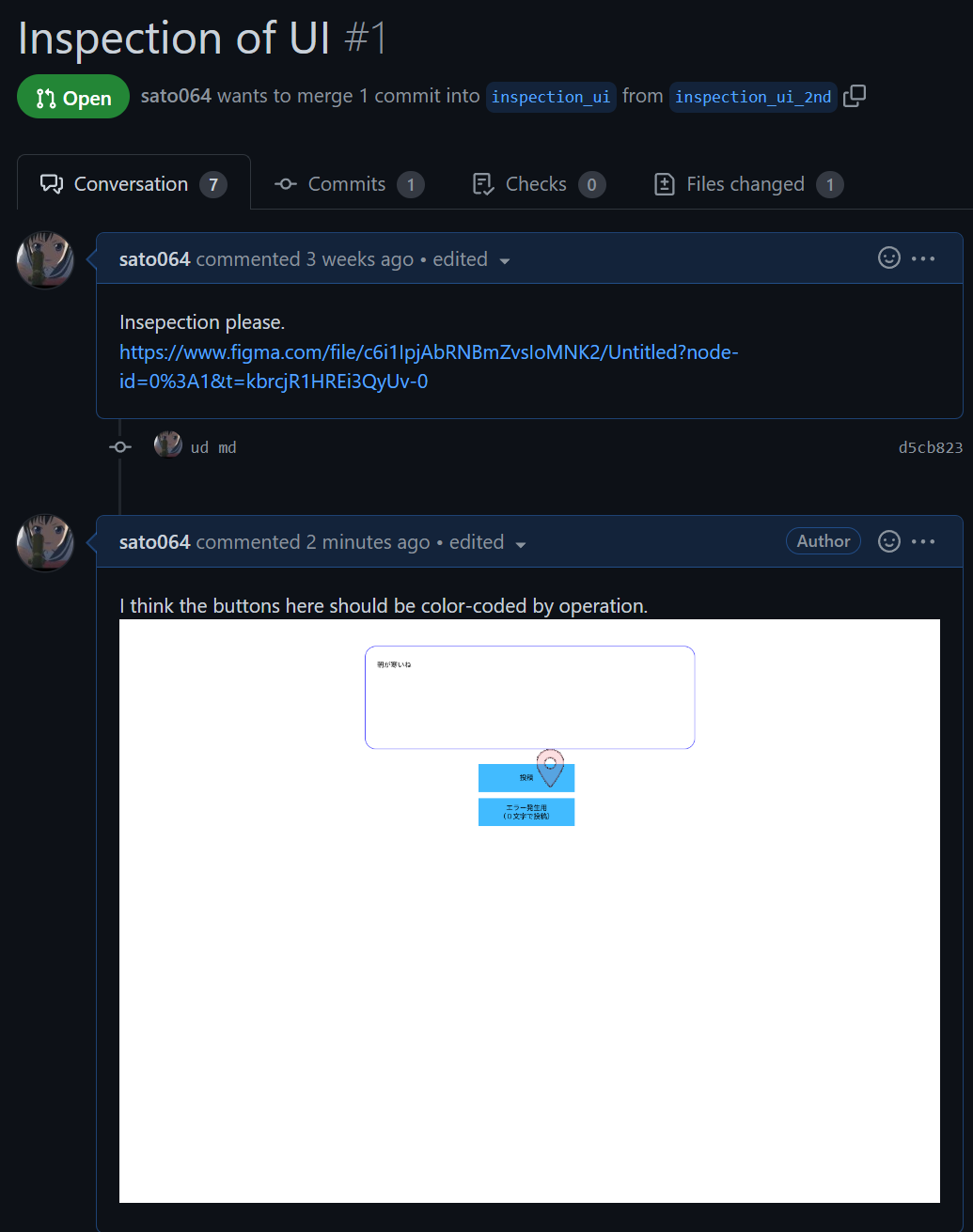}
  \caption{GitHub repository after SceneCommenter works}
\end{figure}

\section{Analysis of Comments}
This section describes the policy for analyzing comments and the results.
\subsection{Target academic year}
This study focuses on three years’ PBL operations during 2020 to 2022. These operations followed the inspection process \cite{miyashita} and the tasks given to the teams were application development for actual educational purposes.  Table IV shows the tasks and clients in three years. 
\begin{table}
  \centering
    \caption{Year, subject and clients}
    \begin{tabular}{|c||p{4.8cm}|c|}  \hline
      Year & Subject & Clients \\ \hline \hline
      2022 & Graduate Division Business Digitization Application & University staff \\ \hline
      2021 & Seating and grade management applications for school classes & Affiliated school teacher \\ \hline
      2020 & Nutrition Education Application for Home Economics Education & University faculty \\ \hline
  \end{tabular}
\end{table}
\begin{table}
  \centering
    \caption{Number of participants and groups in the year in which comments were analyzed}
    \begin{tabular}{|c||c|c|}  \hline
      Year & Number of Student & Number of group \\ \hline \hline
      2022 & 5 & 1 \\ \hline
      2021 & 4 & 1 \\ \hline
      2020 & 10 & 2 \\ \hline
  \end{tabular}
\end{table}
\subsection{Number of comments}
The results of the three-year survey of the number of inspection comments showed that 264 comments were recorded in FY2022, 117 in FY2021, and 171 in FY2020.
Table V shows the number of inspection comments by classification criteria for the past three years, and Figure 4 shows the percentage. Note that the total number of comments by category exceeds the number of inspection comments, but this is due to a single comment may be classified into multiple categories.
\begin{table*}
  \centering
    \caption{Number of inspection comments by classification for the past 3 years}
    \scalebox{0.7}{
    \begin{tabular}{|c||c|c|c|c|c|c|c|c|c|c|c|c|c|}  \hline
      Year,Group & short description & excess & abstract & understandability & undefined & inconsistent & mistake & rationale & short items & missed inspection comments & presentation & enhancement request & format  \\ \hline \hline
      2022G1 & 50 & 13 & 22 & 6 & 3 & 10 & 24 & 22 & 1 & 2 & 57 & 53 & 6  \\ \hline
      2021G1 & 3 & 2 & 8 & 6 & 0 & 8 & 22 & 18 & 2 & 0 & 21 & 33 & 0 \\ \hline
      2020G1 & 14 & 0 & 8 & 8 & 0 & 11 & 21 & 13 & 2 & 1 & 11 & 19 & 2 \\ \hline
      2020G2 & 14 & 1 & 2 & 3 & 2 & 6 & 12 & 6 & 0 & 0 & 12 & 27 & 0 \\ \hline
  \end{tabular}
    }
\end{table*}
\begin{figure*}
  \centering
  \includegraphics[width=\linewidth]{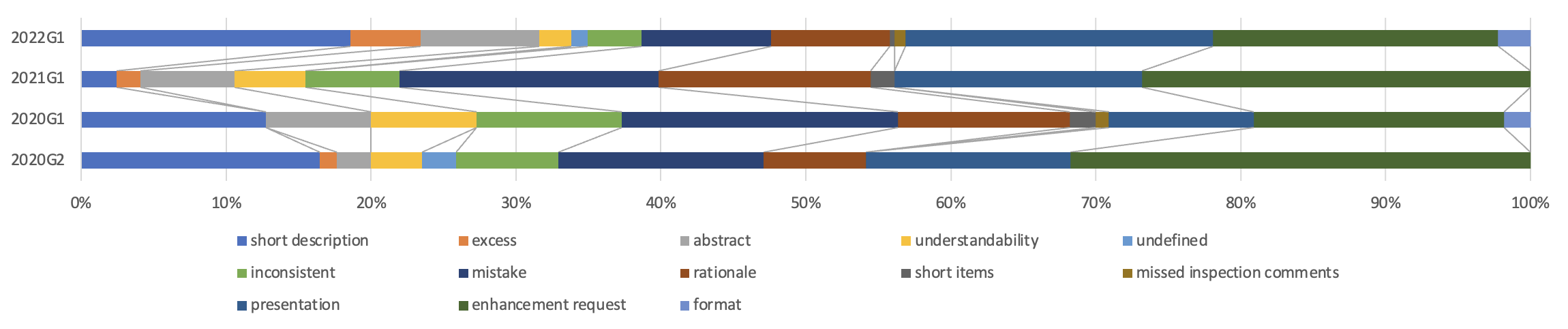}
  \caption{Percentage of inspection comments classified in the past 3 years}
\end{figure*}
\subsection{Analysis Results}
The items that are common to all years and for which a certain percentage exists are abstract, inconsistent, mistakes, rationale, presentation, and suggestions for improvement. In particular, presentation is a simple mistake, yet it is pointed out by about 15\% of the respondents every year. This result suggests the need for a mechanism to promote peer review in order to prevent these kinds of mistakes before they occur.

\section{Relation between student comments and inspection classification}
This section discusses what students learn from the results of the classification of the inspection comments, based on the descriptions of the reports that students submit at the end of the class period.
\par
First, we focus on the comment “I felt that the remarks we received in the inspection were an opportunity for us to show our carelessness and compromise in a big way” from the viewpoint of short description and abstract. The idea that it is sufficient to write this much, or to make vague descriptions because the details are not clear, is considered to appear in these two perspectives.
\par
Next, the perspectives of undefined, inconsistent, mistakes, rationale, and presentation are reflected in the comment "I learned a lot by being pointed out points that I was not aware of myself" by most of the students. These classifications are basically pointing out simple mistakes, and we believe that the students were not aware of these mistakes, even though they were once peer-reviewed. In PBL, students' awareness directly leads to learning, and these remarks directly promote students' learning.
\par
Finally, an interesting comment in terms of enhancement requests. One commenter said, “Inspections and feedback guided the team's development”. First, we can see that the students perceive the inspections as feedback. The students themselves were aware of the meaning of inspections other than the original quality improvement activities. In the aspect of enhancement requests, our enhancement requests may have had a significant impact on the policy decisions of the students' teams.

\section{Future Outlook}
In this section, we discuss prospects based on the above results.

\subsection{Relation between comment categorization and student learning}
In this study, we investigated the impact of inspection comments on students in a software development PBL practiced at a university. The results of categorizing the inspection comments into 13 types clearly identified the items that we consider having been effective in the students' comments. However, some classification items in this study could not be fully measured in terms of their impact on students, and future surveys are expected to be conducted by means of detailed questionnaires to students.
\subsection{Potential utilization of classification }
For the student, it is possible to identify trends in the documents produced by the team by clearly indicating the trends of the observations made during the inspection in the form of a dashboard or other format. This information will make it possible to point out new perspectives in peer review, not only from the students' perspective but also from each other's perspective.
\subsubsection{Student Support}
For the student, it is possible to identify trends in the documents produced by the team by clearly indicating the trends of the observations made during the inspection in the form of a dashboard or other format. This information will make it possible to point out new perspectives in peer review, not only from the students' perspective but also from each other's perspective.
\subsubsection{Teacher Support}
For teachers, it would be possible to determine an optimized teaching strategy by knowing the status of the documents produced by the team. For example, teams that make many errors in presentation are encouraged to check their spelling again, teams that make many inconsistent errors are encouraged to check their past documents, and so on. This support enables students to receive instruction tailored to their characteristics, which is beneficial to both parties.
\subsection{Automatic classification by machine learning}
As described above, classification of software development PBL inspection comments provides various benefits. However, currently it is not possible to automatically classify comments, and manual comment classification takes a considerable amount of time. Therefore, we would like to propose a method to automatically classify inspection comments by machine learning.
\par
Research targeting natural language classification by machine learning related to applications includes the work of Maalej et al\cite{mal}. and Yamada et al\cite{yamaday}. In these studies, application reviews are classified by machine learning. As a result of the script development in this study, inspection comments of software development PBLs can be managed in a text-based manner. By classifying these texts based on the methods of Maalaj et al. and Yamada et al. we believe that the method described in section VII.B can be realized.
\par
In order to implement the advantages described in section VII.B in PBL activities and measure their effectiveness, we are planning to develop an application that supports software development PBL by automatically classifying comments using machine learning and automatic classification.
\section{Conclusion}
The purpose of this study was to classify the inspection comments made in a software development PBL and to investigate their impact on students. The comments were compiled from the past three years of practice, and the impact on students was discussed based on the students' post-lesson questionnaires.
\par
During our research, we also developed a script that can handle comments from different platforms (GitHub and Figma) in a centralized text-based manner. In addition, we summarized the benefits of comment categorization for students and professors, and proposed a new PBL support that enables automatic categorization by machine learning.
\par
In the future, we would like to practice automatic classification by machine learning and develop an application to support software development PBL through comment classification.

\end{document}